\documentclass[structabstract]{aa}  
\usepackage{graphicx}
\usepackage{lscape}
\usepackage{txfonts}
\usepackage{natbib}
\defcitealias{Taka08}{T08}
\begin{document}
   \title{Warm gas in protostellar outflows}

   \subtitle{II. EHV jet and outflows from OMC-2/3 \thanks{Based on observations acquired with the Atacama Pathfinder Experiment (APEX). APEX is a collaboration between the Max-Planck-Institut f\"ur Radioastronomie, the European Southern Observatory, and the Onsala Space Observatory.}}

   \author{A. I. G\'omez-Ruiz
          \inst{1},
           A. Gusdorf\inst{2}, S. Leurini\inst{3}, K. M. Menten\inst{4}, S. Takahashi\inst{5,6,7}, F. Wyrowski\inst{4}, R. G\"usten\inst{4}
          }
     \institute{CONACYT--Instituto Nacional de Astrof\'isica, \'Optica y Electr\'onica, Luis E. Erro 1, C.P. 72840, Tonantzintla, Puebla, M\'exico \\
\email{aigomez@inaoep.mx}
\and
             LERMA, UMR 8112 du CNRS, Observatoire de Paris, \'Ecole Normale Sup\'erieure, 61 Av. de l'Observatoire, 75014, Paris, France
	\and
INAF-Osservatorio Astronomico di Cagliari, Via della Scienza 5, I-09047, Selargius (CA), Italy
       \and
   Max-Planck-Institut f\"ur Radioastronomie (MPIfR),
              Auf dem H\"ugel 69, 53121 Bonn, Germany
         \and
Joint ALMA Observatory, Alonso de C\'ordova 3107, Vitacura, Santiago, Chile
             \and
NAOJ Chile Observatory, Alonso de C{\'{o}}rdova 3788, Oficina 61B, Vitacura, Santiago, Chile, 
\and
Department of Astronomical Science, School of Physical Sciences, SOKENDAI (The Graduate University for Advanced Studies), Mitaka, Tokyo 181-8588, Japan
}
   \date{Received ; accepted }

 
  \abstract
   {OMC-2/3 is one of the nearest embedded cluster-forming region that includes intermediate-mass protostars at early stages of evolution. A previous CO (3--2) mapping survey towards this region revealed outflow activity related to sources at different evolutionary phases.}
   {The present work aims to study the warm gas in the high-velocity emission from several outflows found in CO (3--2) emission by previous observations, determine their physical conditions, and make comparison with previous results in low-mass star-forming regions.}
   {We use the CHAMP+ heterodyne array on the APEX telescope to map the CO (6--5) and CO (7--6) emission in the OMC-2 FIR 6 and OMC-3 MMS 1-6 regions, and to observe $^{13}$CO (6--5) at selected positions. We analyze these data together with previous CO (3--2) observations. In addition, we mapped the SiO (5--4) emission in OMC-2 FIR 6.}
   {The CO (6--5) emission was detected in most of the outflow lobes in the mapped regions, while the CO (7--6) was found mostly from the OMC-3 outflows. In the OMC-3 MMS 5 outflow, a previously undetected extremely high velocity gas was found in CO (6--5). This extremely high velocity emission arises from the regions close to the central object MMS 5. Radiative transfer models revealed that the high-velocity gas from MMS 5 outflow consists of gas with $n_{\rm H_2}=$10$^4$--10$^{5}$ cm$^{-3}$ and $T>$ 200 K, similar to what is observed in young Class 0 low-mass protostars. For the other outflows, values of $n_{\rm H_2}>$10$^{4}$ cm$^{-3}$ were found.}
   {The physical conditions and kinematic properties of the young intermediate-mass outflows presented here are similar to those found in outflows from Class 0 low-mass objects. Due to their excitation requirements, mid$-J$ CO lines are good tracers of extremely high velocity gas in young outflows likely related to jets.}

\titlerunning{Warm gas in outflows II}
\authorrunning{G\'omez-Ruiz, Gusdorf, Leurini et al.}

   \maketitle
%

\section{Introduction}

Intermediate-Mass (IM) protostars, i.e. protostars whose mass is in the range from 2 to 8 $M_\odot$, provide crucial information on the star formation process because they link the low- and high-mass star formation \citep[e.g.][]{Crimer10}. IM protostars share some properties of massive protostars, such as their existence in clusters and their ability to ionize the surrounding gas. On the other hand, they present the advantage that they are more numerous and therefore located at closer distances ($\lesssim$1 kpc) than most of the massive star-forming regions. Therefore they could fill the gap between low- and high-mass star formation studies. Despite their relevance, very little is known about the formation and early evolutionary stages of IM protostars (e.g. class 0 type objects). Although some detailed studies have been conducted in the last years \citep[e.g.,][]{Beltran15,Taka12,Taka09,Taka06,Kempen12,Crimer10,Beltran08,Fuente07,Fuente01,Zapata05,Zapata04}, little information is available, in particular, about outflows from class 0 type sources.

Observations have shown that outflows from IM protostars (hereafter IM outflows) share some characteristics of outflows from low-mass protostars, such as the high collimation factor in early evolutionary phases \citep{Beltran08}. On the other hand, in general, IM outflows are intrinsically more energetic than those driven by low-mass sources \citep{Beltran08,Taka08}. The values of the outflow momentum rate, $F_{\rm out}$, are higher in IM protostars than in low-mass objects, which may be an indication that IM young stellar objects (YSOs) accrete material faster (higher accretion rate) than the low-mass ones \citep{Beltran08}. 

Regarding their physical conditions, such as kinetic temperature (T$_{\rm kin}$) and H$_2$ density (n), only for a few cases these have been determined via observations of more than one CO transition \citep[e.g., Cep-E and NGC2071:][]{GR12,Chernin92,Hatchell99}. The best studied case is Cep-E, in which a multi-line CO study at high spectral resolution uncovered different excitation components related to the different kinematic structures \citep{Lefloch15}.  In the jet component, the emission from the low-J CO lines (from J=1--0 to J=5--4) is dominated by a gas layer with T$_{\rm kin}$ = 80--100 K, and n $=$ (0.5--1) $\times$ 10$^5$ cm$^{-3}$; while the high-J CO lines (from J=12--11 to J=16--15) trace warmer and denser gas, with T$_{\rm kin}$ = 400--750 K, and n = (0.5--1) $\times$ 10$^6$ cm$^{-3}$. In the outflow cavity, the low-J CO lines are dominated by a gas layer with T$_{\rm kin}$ = 55--85 K, and density in the range (1--8) $\times$ 10$^5$ cm$^{-3}$; while the high-J lines are dominated by a hot, denser gas layer with T$_{\rm kin}$ = 500--1500 K, and n = (1--5) $\times$ 10$^6$ cm$^{-3}$. The terminal bowshock consists of gas with moderate excitation, with a temperature in the range T$_{\rm kin}$ $=$ 400--500 K, and density n $=$ (1--2) $\times$ 10$^6$ cm$^{-3}$. These observations showed a complex excitation structure of the outflow, with the CO (5--4) line (close to what we later call mid-J CO lines) contributing mostly to a relatively low-excitation component from the jet and cavity, and a moderate excitation component from the bowshock.


In a previous paper \citep[][hereafter Paper I]{GR13}, we have presented a study of the CO (6--5) and (7--6) emission (throughout the paper referred to as mid$-J$ CO transitions) in two low-mass class 0 outflows, namely L1448 and HH211. In both objects we found CO (6--5) and (7--6) emission tracing bipolar structures. In the case of L1448, extremely high velocity emission (EHV), i.e. emission at relative velocities $>$50 km s$^{-1}$ with respect to the cloud velocity, was detected in both transitions. By performing a large velocity gradient (LVG) analysis, we inferred that the gas related with the outflow structures is dense ($>$10$^{5}$ cm$^{-3}$) and warm (T$>$ 200 K). It was determined that the EHV emission shows higher densities and temperatures than the high-velocity emission. Based on the LVG results and the position-velocity distribution of the emitting gas, we concluded that these mid$-J$ CO transitions are good tracers of molecular material likely related to the primary jet and to the highly excited bow-shock region in protostellar outflows.

In the present paper we extend the study of mid$-J$ CO emission to outflows from IM protostars in the Orion molecular cloud-2/3 (OMC-2/3), including Class 0 and I type objects. The main purpose of this paper is to study the distribution of the warm gas traced by the CO (6--5) and (7--6) transitions in IM outflows and compare it with the low-mass case. By incorporating lower$-J$ CO transitions in our analysis, the excitation and physical conditions of IM outflows, as traced by these lines, are also studied. 

   \begin{figure}
   \centering
   \includegraphics[bb=52 75 556 699,width=8.5cm,angle=0]{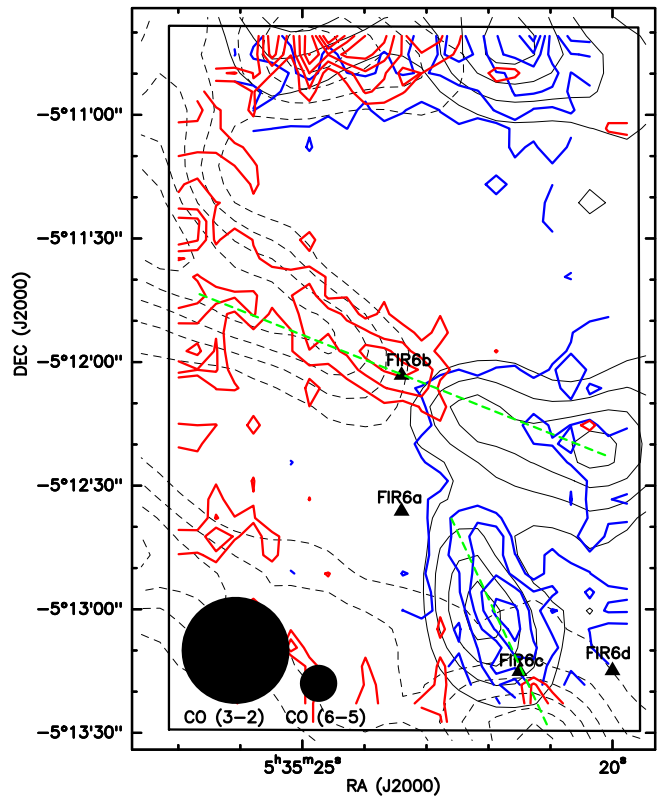}
\caption{The CO (6--5) outflow emission in the OMC-2 FIR 6 region. Blue contours are for the blue-shifted emission in the range from -12 to $+$8 km s$^{-1}$; while red contours are for the red-shifted emission from $+$14 to $+$30 km s$^{-1}$. First contour is 3$\sigma$, with contour spacing of 2$\sigma$ ($\sigma =$5.5 and 2.4 K km s$^{-1}$, for blue- and red-shifted, respectively). The black solid and dashed contours are the blue- and red-shifted CO (3--2) emission, respectively, as presented by T08. The black square shows the whole field covered by our CHAMP+ observations. The triangles show the positions of the continuum sources and the green dashed indicate the outflows orientation (both according to T08).
              }
         \label{fir6-co65-wings}
   \end{figure}

\section{The target regions}

   \begin{figure*}
   \includegraphics[bb=517 72 166 716,width=9cm,angle=90]{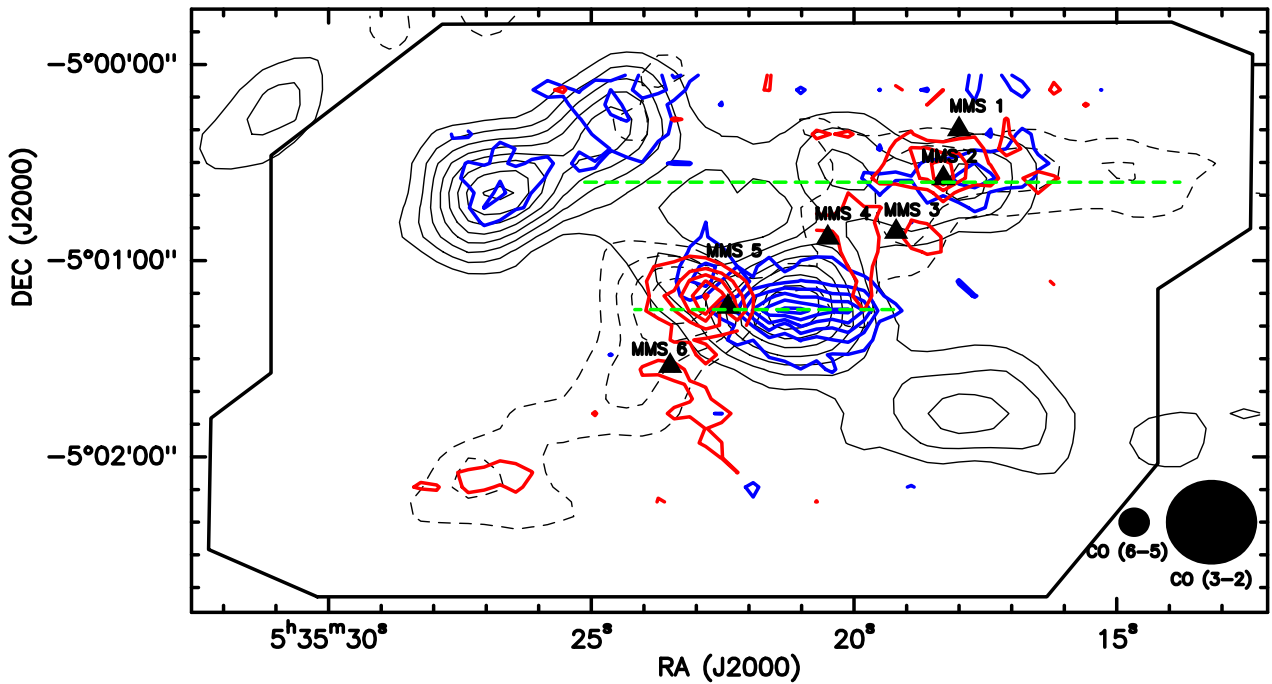}
      \caption{The CO (6--5) outflow emission in the OMC-3 MMS 1-6 region. Blue contours are for the blue-shifted emission in the range from -25 to $+$7 km s$^{-1}$; while red contours are for the red-shifted emission from $+$16 to $+$31 km s$^{-1}$. First contour is 3$\sigma$ and contour spacing is 2$\sigma$ ($\sigma =$1.9 and 0.8 K km s$^{-1}$, for blue- and red-shifted, respectively). The black solid and dashed contours are the blue- and red-shifted CO (3--2) emission, respectively, as presented by T08. The black square shows the whole field covered by our CHAMP+ observations. The triangles show the positions of the continuum sources and the green dashed indicate the outflows orientation (both according to T08).
              }
         \label{mms5-co65-wings}
   \end{figure*}

OMC-2/3 is located north of the Orion-A Giant Molecular Cloud (or OMC 1). The distance to OMC 1 range from $\sim$ 390pc to 440 pc, given its 3-D structure determined by recent astrometry data \citep{Kounkel18}. This region is one of the nearest cluster-forming sites with intermediate-mass protostars \cite[e.g.][here after T08]{Taka06,Taka08}. In this region, filamentary molecular clouds were discovered in the millimeter/sub-millimeter continuum observations with the IRAM-30m telescope and the JCMT \cite[]{Chini97,Johns99}. Within the filaments, millimeter/sub-millimeter continuum sources are found, presumably tracing prestellar and protostellar cores \cite[e.g.,][]{Taka09,Taka12b,Taka12,Taka13,Chini97}. Follow-up CO and H$_2$ molecular line observations revealed the presence of a number of outflows originated from some of the continuum sources \citep[T08;][]{Yu97,Aso00,Stanke02,Will03}. One of the most extensive molecular outflow surveys in the OMC-2/3 region has been conducted by T08. By observing the CO (3--2) transition with the ASTE telescope, these authors detected 14 molecular outflows, indicative of a high ongoing star formation activity in this region. 

From the CO (3--2) outflow survey of T08 and its follow-up observations, we selected the most promising sources in terms of outflow activity (bipolar morphology and high-velocity emission) and early evolutionary stage (class 0 type sources). Under these criteria, the regions around FIR 6 in OMC-2 and MMS 1-6 in OMC-3 (hereafter OMC-2 FIR 6 and OMC-3 MMS 1-6, respectively) were selected for follow-up observations in mid$-J$ CO lines. For these sources the derived luminosities are in the range of 6--251 and 9--150 $L_{\odot}$, respectively, which makes them very likely IM protostars \citep[T08,][]{Taka09,Furlan16}. The envelope masses determined by T08 ($>$ 3 M$_{\odot}$) are consistent with values of IM protostars. 

OMC-2 FIR 6 contains at least four millimeter continuum sources named FIR 6a, 6b, 6c, and 6d (see Figure \ref{fir6-co65-wings}). T08 reported the FIR 6b source as being likely a Class I type object (based on their SED) while the lack of sufficient information for the rest of the sources prevents to clearly assess their evolutionary status. The CO (3--2) data towards OMC-2 FIR 6 region revealed that two molecular outflows are driven by FIR 6b and 6c (see Figure \ref{fir6-co65-wings}). As a follow-up study of T08, \citet{Shima09} observed the OMC-2 FIR 6 region with the Nobeyama Millimeter Array in the 3 mm continuum, CO (1--0), and SiO (2--1) transitions. The two outflows found in CO (3--2) were confirmed in CO (1--0), in addition to a third outflow emanating from FIR 6d. The SiO (2--1) emission was found only at the blue-lobe of FIR 6c outflow, but no SiO emission was clearly detected towards the 6b outflow. From the line ratio between the CO (3--2) and (1--0) transitions, it was found that kinetic temperature in the outflow lobes is $\sim$50 K \citep[T08;][]{Shima09}. 

The OMC-3 MMS 1-6 region consists of six millimeter continuum sources, named by \citet{Chini97} as MMS (see Figure \ref{mms5-co65-wings}). A clear identification of the evolutionary status through the SED was given for MMS 2 and MMS 5, which are class I and class 0 type, respectively \citep[T08;][]{Taka09}. The MMS 2 source is actually a binary system of class I sources \citep[T08;][]{Tsuji04}. T08 reported clear bipolar outflow morphologies originating from MMS 2 and MMS 5 in the CO (3-2) emission. In the case of MMS 6, sub-arcsec resolution observations with the SMA in the CO (3--2) line by \citet{Taka12} revealed an extremely compact molecular outflow (lobe size of $\approx$ 800 AU) associated with the continuum peak named MMS 6-main. Furthermore, Takahashi et al. (2012) reported that MMS 6-main is most likely in the protostellar phase. No clear molecular outflow signatures are detected in the other MMS sources, so that they are likely in the prestellar phase \citep{Taka13}.

\section{APEX observations}

\begin{table}
\begin{minipage}{\columnwidth}
\caption{Map central position and r.m.s.}             
\label{sources}
\renewcommand{\footnoterule}{} 
\centering          
\begin{tabular}{l c c c c c}     
\hline       
\hline
       &           &            & \multicolumn{3}{c}{r.m.s. (K)\footnote{Average r.m.s. ($T_{\rm MB}$) in a channel map of 1.0 km s$^{-1}$.}}\\
\cline{4-6}
Source & $\alpha$(J2000) & $\delta$(J2000) & \multicolumn{2}{c}{$^{12}$CO} & SiO  \\
       & ($^{\rm h}$ $^{\rm m}$ $^{\rm s}$)     & ($^{\circ}$ $'$ $''$) & (6--5) & (7--6) & (5--4) \\
\hline                    
OMC-2 FIR 6b   &05:35:23.4 & $-$05:12:03    & { 0.7} & 3.0 & 0.05 \\
OMC-3 MMS 5   &05:35:22.5 & $-$05:01:15	& { 0.5} & 1.9 & -	\\
\hline         
\end{tabular}
\end{minipage}
\end{table}

\begin{table}
\caption{Characteristics of the observed transitions.}             
\label{setup}      
\centering          
\begin{tabular}{l c c c c}     
\hline       
Line   & $\nu$$_0$   & $A$ & E$_u$/k   & HPBW   \\
       &  GHz        & 10$^{-5}$ s$^{-1}$ &    K      & $''$\\
\hline
CO (6--5)         &  691.473  & 2.137& 116.16 & 9.0    \\
CO (7--6)         &  806.651  & 3.422& 154.87 & 7.7    \\
$^{13}$CO (6--5)   & 661.067   & 299.0& 111.05 & 9.4    \\
SiO (5--4)        & 217.107   & 51.96& 31.26  & 30.5   \\
HCO$^+$ (9--8)    & 802.458   & 4400 & 192.58 & 7.8    \\
\hline 
\multicolumn{5}{c}{Note--Molecular data taken from LAMDA \citep{LAMDA}.}\\           
\end{tabular}
\end{table}

\subsection{CHAMP+ observations}

Submillimeter observations towards OMC-3 MMS 1-6 and OMC-2 FIR 6 were performed with the MPIfR principal investigator (PI) instrument CHAMP$^+$ \citep{Kase06} on the APEX telescope \citep{Rolf06} during November 2008 and July 2010, respectively.  CHAMP$^+$ is a dual-color 2 $\times$ 7 pixels heterodyne array for operation in the 450 $\mu$m (low frequency array, LFA) and 350 $\mu$m (high frequency array, HFA) atmospheric windows. The optics allow simultaneous observations in both colors. Both sub-arrays present an hexagonal arrangement. The front-end was connected to a Fast Fourier Transform Spectrometer \citep[FFTS,][]{Klein06}. 

The CO (6--5) and (7--6) lines were mapped simultaneously in the On-The-Fly (OTF) mode, with an ON time of 0.5 second per position, with steps of 3$''$. For the OMC-2 FIR 6 region a map of 60$'' \times$120 $''$ was centered at the position of the FIR 6b source. For the OMC-3 MMS 1-6 region a map with of 280$''\times$150$''$ was centered at the position of MMS 5. Additional single-pointing observations were done on two selected positions of OMC-2 FIR 6 (called FIR 6b and FIR 6c-B1; coordinates are listed in Table \ref{sources} and \ref{positions}, respectively) in a setup consisting of $^{13}$CO (6--5) and HCO$^+$ (9--8). No emission was detected in the HCO$^+$ (9--8) line, with upper limits of 186 mK ($T_{\rm A}^*$), and therefore not further discussed. For all observations, the backend was set to provide a total bandwidth of 2.8 GHz divided in 8192 channels for each pixel. The final spectra were re-sampled to 1.0 km s$^{-1}$ spectral resolution for all transitions in both sources.

During the observations, the precipitable water vapor (PWV) was in the range of 0.3 to 0.7 mm. The pointing was determined by CO (6--5) cross-scan observations on IK Tau. Pointing accuracy was always within 3$''$. Focus was checked on Jupiter and Mars. The calibration was done by observing hot and cold loads. In our observations the OFF position was not checked for emission, but this position was selected to avoid regions known to present emission in low$-J$ CO transitions. However, due to a erroneous selection of the OFF position during the $^{13}$CO(6--5) and HCO$^+$(9--8) single-pointing observations of the FIR 6b position, contamination of the OFF position was found in these spectra. Standard data reduction procedures such as flagging bad spectra and base line subtraction were done with CLASS, while images were produced with the XY$\_$MAP task in GReG, both programs being part of the GILDAS software \footnote{http://iram.fr/IRAMFR/PDB/gildas/gildas.html}. The griding procedure in XY$\_$map was made to provide a final resolution equal to the HPBW of the telescope at the observed frequency. Table \ref{setup} lists the observed transitions and rest frequencies ($\nu _0$), their corresponding HPBW, upper state energy ($E_u$) and Einstein coefficient ($A$). Table \ref{sources} shows the coordinates (right ascension and declination) of the central position of the maps and their average r.m.s.   

The images and spectra are presented in main beam brightness temperature scale, $T_{\rm MB} = T_{\rm A}^*$/$\eta_{\rm MB}$ ($\eta_{\rm MB}=\eta_{\rm s}/F_{\rm eff}$), for which we have used a forward efficiency ($F_{\rm eff}$) of 0.95 and a beam coupling efficiency ($\eta_{\rm s}$) measured on Jupiter. During the OMC-2 observations the measured $\eta_{\rm s}$ was 0.48 for both LFA and HFA, while during the OMC-3 observations $\eta_{\rm s}$ was 0.48 and 0.45 for LFA and HFA, respectively. Based on the variation of the integrated mid-J CO flux observed in the line pointing sources, we estimate 20 \% calibration uncertainty for both, LFA and HFA \citep[see also][]{Yildiz15}.  The r.m.s. noise levels are indicated in Table \ref{sources}.


%

\subsection{APEX-1 observations}

Complementary mapping observations in the SiO (5--4) line were carried out towards OMC-2 FIR 6 with the APEX-1 facility instrument in September 2010. The receiver was connected to a FFTS that provides a bandwidth of 1 GHz and 4096 channels,  which at the rest frequency of SiO (5--4) resulted in a spectral resolution of 0.3 km s$^{-1}$. The OTF map with a size of 70$''$$\times$70$''$, in steps of 9$''$, was centered at the position of the FIR 6b source. Therefore, the map covers the central parts of the FIR 6b outflow's lobes and the blue lobe of the FIR 6c outflow. The pointing accuracy was checked by CO (2--1) cross-scan observations on IK Tau. Pointing accuracy was always within 5$''$. The focus was checked on Jupiter and Mars. The calibration was done by observing hot and cold loads. The data reduction and image processing are similar to the procedure described for the CHAMP+ observations, using CLASS and GReG. The final spectra were re-sampled to a 2.0 km s$^{-1}$ spectral resolution. In order to convert to $T_{\rm MB}$ scale, the nominal values of 0.95 and 0.75 were adopted for the forward and beam coupling efficiency, respectively. A calibration uncertainly of 20 \% is also expected for this receiver, based on the flux variation of the line pointing sources. 

\begin{table}
\caption{Coordinates of the outflow peak positions.}             
\label{positions}      
\centering          
\begin{tabular}{l c c c}     
\hline       
\hline
Position & $\alpha$(J2000) & $\delta$(J2000) & $\varv_{\rm min}$,$\varv_{\rm max}$*  \\
       & ($^{\rm h}$ $^{\rm m}$ $^{\rm s}$)     & ($^{\circ}$ $'$ $''$) & (km s$^{-1}$) \\
\hline                    
FIR 6b-B1   &05:35:21.2 & $-$05:12:15    & 5,8 \\
FIR 6b-R1   &05:35:23.7 & $-$05:11:59    & 14,30 \\
FIR 6c-B1   &05:35:21.6 & $-$05:13:00    & -16,8  \\
FIR 6c-B2   &05:35:21.7 & $-$05:13:00    & -10,8  \\
MMS 5-B1    &05:35:21.3 & $-$05:01:16    & -95,7  \\
MMS 5-R1    &05:35:22.5 & $-$05:01:14    & 16,75  \\
MMS 2-B1    &05:35:18.2 & $-$05:00:32    & -12,7  \\
MMS 2-R1    &05:35:17.7 & $-$05:00:34    & 16,30   \\ 
\hline    
\multicolumn{4}{c}{*Velocity boundaries of the outflow emission (see Sect. 4).}\\     
\end{tabular}
\end{table}

\section{Results}

   \begin{figure}
   \centering
   \includegraphics[bb=74 199 340 716,width=6cm,angle=0]{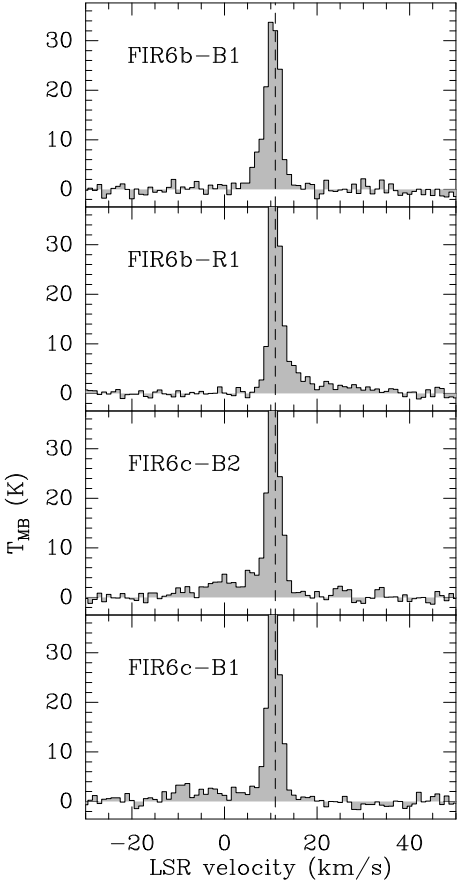}
      \caption{CO (6--5) sample spectra at selected positions (see Table 3) within FIR 6 outflows (1 km s$^{-1}$ spectral resolution). Dashed line indicates the cloud velocity.
              }
         \label{spect-65-fir6}
   \end{figure}

   \begin{figure*}
   \centering
   \hbox{
   \includegraphics[width=6.6cm,angle=-90]{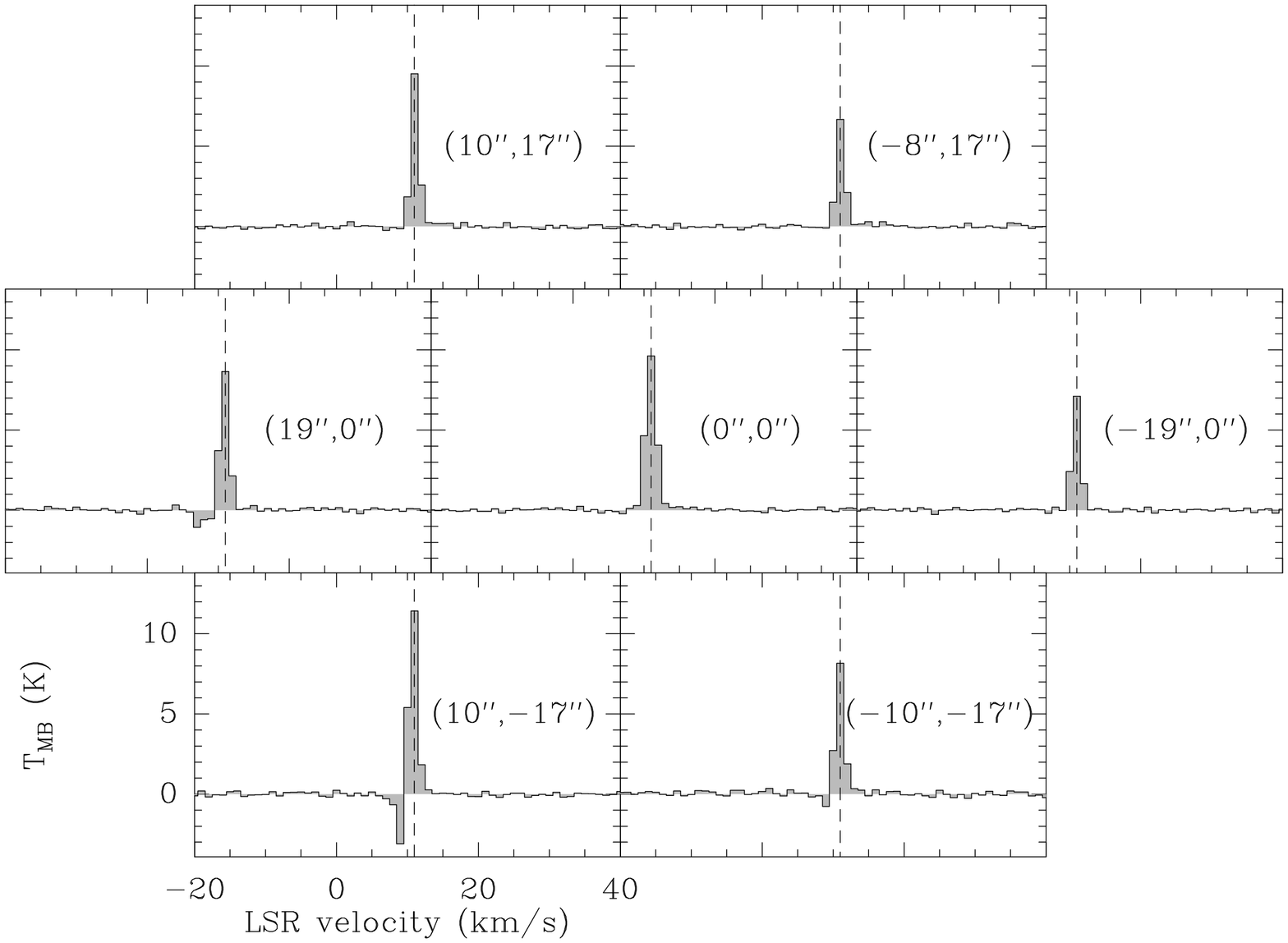}
   \includegraphics[width=6.6cm,angle=-90]{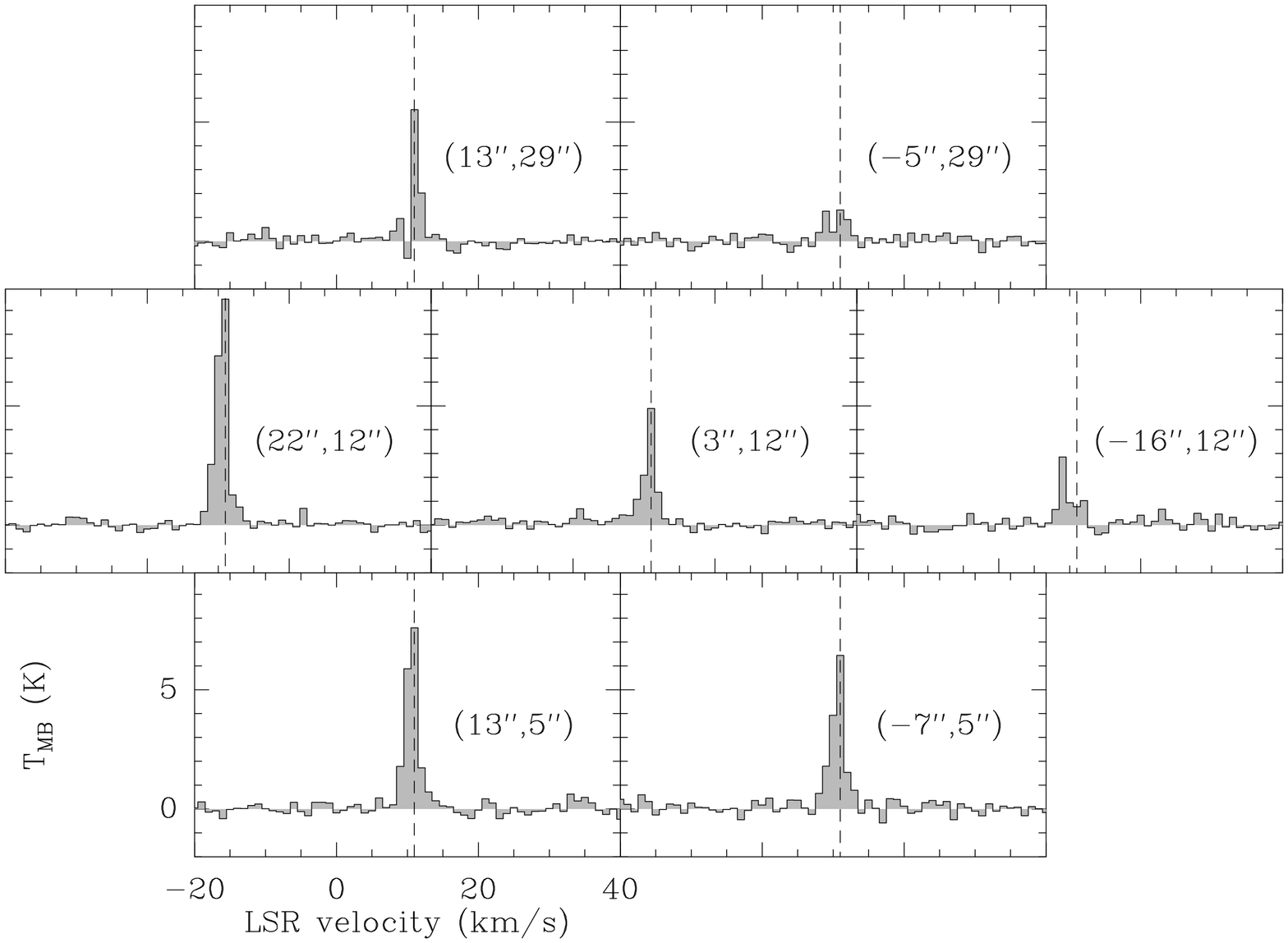} }
      \caption{$^{13}$CO (6--5) spectra taken at selected positions within FIR 6 region (1 km s$^{-1}$ spectral resolution). The figure shows the seven CHAMP+ pixels for each pointing observation. {\it Left:} observation centered on FIR 6b. {\it Right:} observation centered on FIR 6c-B1 position (offsets with respect to FIR 6c).  
              }
         \label{spect-1365-fir6}
   \end{figure*}


In order to define the outflow velocity range we use the CO (6--5) channel maps and the spectra taken at different outflow positions. We define the lower boundary limit ($\varv_{min}$) by looking for outflow morphologies clearly separated from the ambient diffuse emission. We then establish $\varv_{max}$ from the spectra taken at the peak positions within the outflow's lobes. In our case, $\varv_{max}$ is defined as the velocity in which the emission in the spectrum drops below the 3$\sigma$ level. The velocity limits for all identified outflows as well as the positions of the outflow peaks are reported in Table \ref{positions}. Based on the measurements of the H$^{13}$CO$^+$ (1--0) line by \citet{Aso00}, throughout the paper we assume the cloud LSR velocity (V$_{\rm LSR}$) of $+$11 km s$^{-1}$. 


\subsection{OMC-2 FIR 6 outflows}

\subsubsection{CO (6--5) and CO (7--6) emission}

Figure \ref{fir6-co65-wings} shows the CO (6--5) emission detected in the FIR 6 region, integrated along the blue- and red-shifted wings. Figure \ref{fir6-co65-wings} also shows the CO (3--2) wing emission from T08. We found the two outflows related to FIR 6b and FIR 6c, reported previously in CO (3--2) by T08. Both outflow lobes from FIR 6b are detected in CO (6--5), and also the blue-shifted lobe from FIR 6c (the red-shifted lobe was not covered by our map). Four emission peaks were found at the outflow positions, which we list in Table \ref{positions}. The CO (6--5) emission from FIR6b seems to peak closer to the central source than the CO (3--2) emission. Convolutions to the CO (3--2) beam confirms that this tendency is not an effect of the different angular resolutions. On the other hand, the peak position of the CO (6--5) emission in the blue lobe of FIR 6c outflow is consistent with the location of the CO (3--2) emission peak. No high-velocity emission tracing outflows from FIR 6a and FIR 6d was found. In the case of FIR 6d we are not able to unambiguously probe the outflow emission, since the outflow structures revealed by previous low$-J$ CO observations are located right at the edge of our map. Around the northern edge of the map, blue- and red-shifted emission is also noticed; however, these structures are not classified as a clear outflow by T08, and therefore we exclude these structures in the analysis that follows.


Figure \ref{spect-65-fir6} shows the spectra taken at the position of the emission peak of each lobe. In all positions, the peak of the CO (6--5) profiles is found at the cloud systemic velocity (V$_{LSR} \approx$11.0 km s$^{-1}$). Three positions show wings (FIR 6b-R1) and secondary peaks at high-velocities (FIR 6c-B1 and FIR 6c-B2). At the FIR 6b-B1 position the line profile only extends down to V$_{LSR}=$5 km s$^{-1}$.

    
The CO (7--6) emission was detected mostly at low velocities (V$_{LSR}=$ 8 to 14 km s$^{-1}$), tracing the cloud/envelope emission. At the outflow peaks the CO (7--6) emission is barely noticed (S/N $<$ 5) in the spectra.

     \begin{figure}
 \vbox{
      \includegraphics[width=8.5cm,angle=-90]{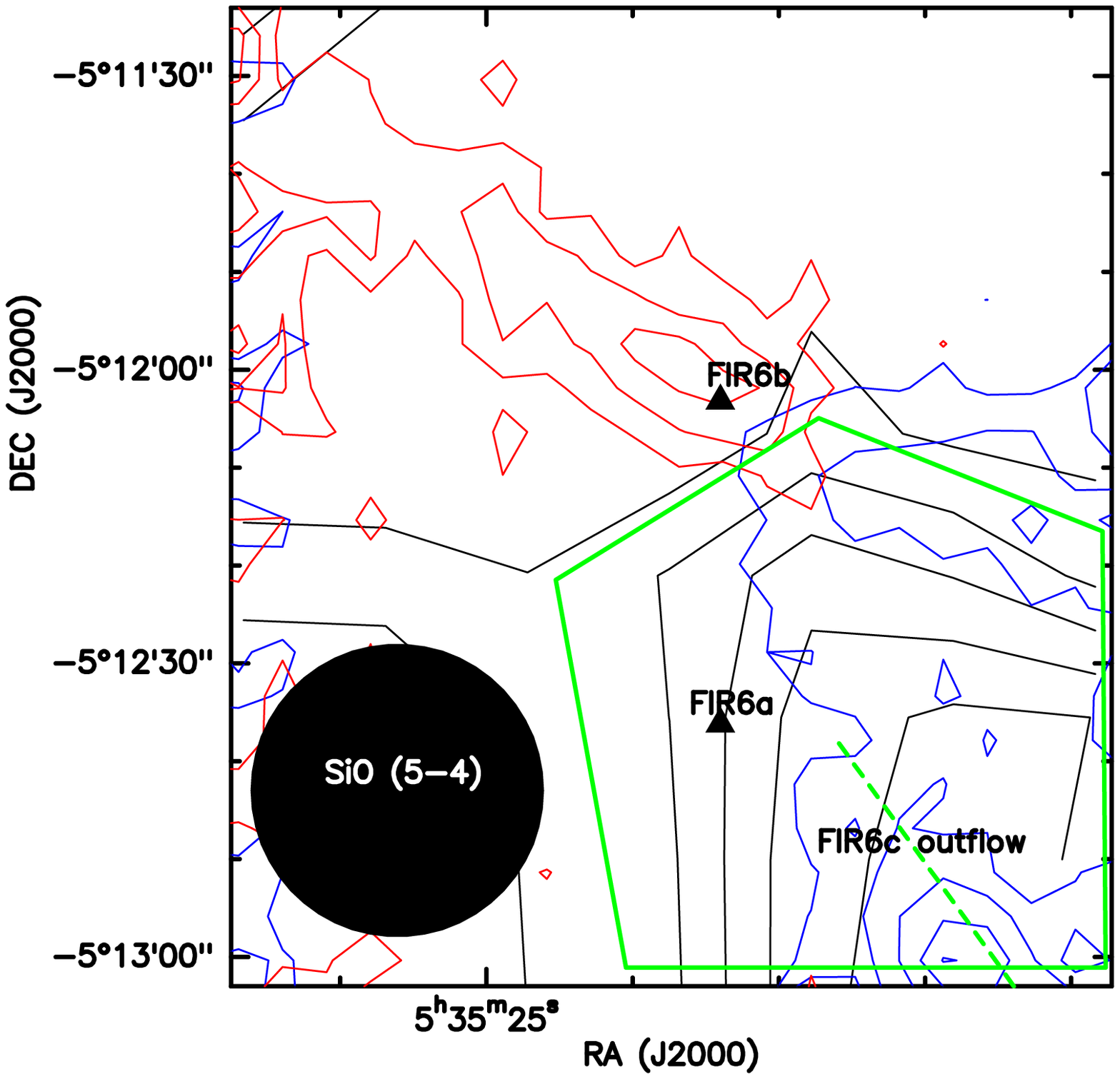}
      \includegraphics[angle=0,width=8cm]{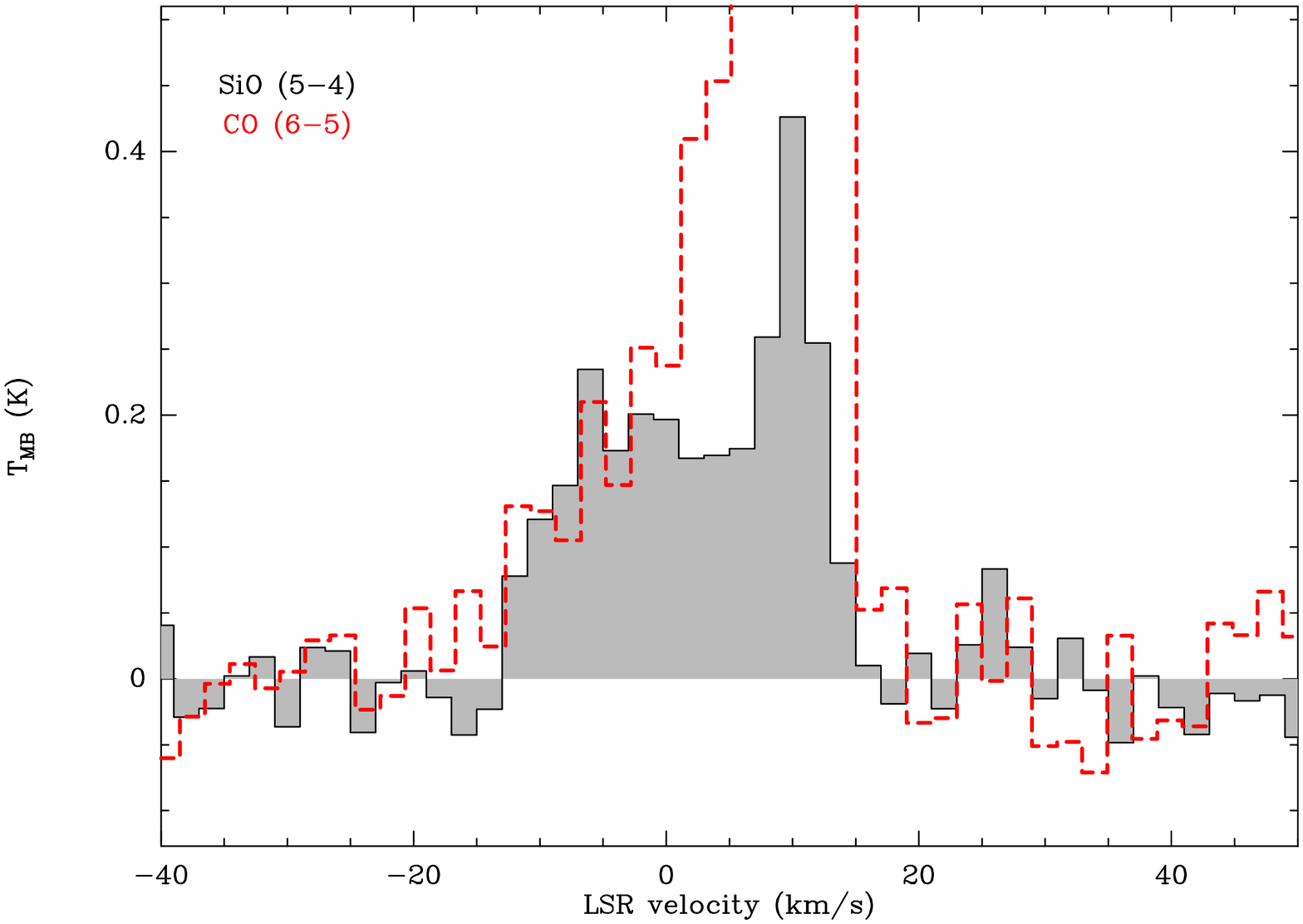}}
      \caption{Upper panel: SiO (5--4) map of the FIR 6 region, integrated from -12 to 15 km s$^{-1}$ (black contours), overlaid with the CO (6--5) outflow emission (blue and red contours; as in Figure \ref{fir6-co65-wings}). SiO contour step is 3$\sigma$, with the first contour at 3$\sigma$ ($\sigma$=0.38 K km s$^{-1}$). Lower panel: SiO (5--4) spectrum (black) averaged over the region indicated by the green polygon on the top panel. To increase the signal-to-noise ratio the spectral resolution has been smoothed to 2 km s$^{-1}$. As comparison the CO (6--5) spectrum is shown in red.  
              }
         \label{siomap}
   \end{figure}

\subsubsection{$^{13}$CO (6--5) emission and opacity estimation}

In Figure \ref{spect-1365-fir6} we show the $^{13}$CO (6--5) spectra taken at two positions within the FIR 6 region: the FIR 6b source and FIR 6c-B1. The $^{13}$CO (6--5) emission is only detected in about three channels, around the cloud velocity. All spectra in Figure \ref{spect-1365-fir6} show the emission peaking at the systemic velocity (V$_{\rm LSR}=$11 km s$^{-1}$), with a FWHM of $\sim$2 km s$^{-1}$. At the FIR 6b position an absorption feature is also present at V$_{\rm LSR}=$9 km s$^{-1}$, which is more evident in the external pixels. This absorption feature is likely due to contamination from the OFF position (see Sect. 2), and thus the $^{13}$CO (6--5) emission detected at the FIR 6b position should be taken as a lower limit.   



\begin{table}
\caption{$^{12}$CO (6--5) opacity around cloud V$_{\rm LSR}$ in FIR 6b and FIR 6c-B1.}             
\label{opacity}      
\centering          
\begin{tabular}{l c c}     
\hline
\hline       
V$_{\rm LSR}$ & $^{12}$CO/$^{13}$CO (6--5) & opacities  \\
  (km s$^{-1}$) &  $T_{\rm MB}$ ratio         & \\
\hline
\multicolumn{3}{c}{FIR 6b}\\
\hline                    
10   & 8.6(0.3)  & 5.8(0.3)\\
11   & 4.8(0.1)  & 10.4(0.2)\\
12   & 7.3(0.3)  & 6.8(0.3)\\
\hline
\multicolumn{3}{c}{FIR 6c-B1}\\                    
\hline
10   & 10.1(0.5)  & 4.9(0.2)\\
11   & 5.5(0.2)  &  9.1(0.3)\\
12   & 18.2(2.0)  & 2.5(0.3)\\ 
\hline  
\hline       
\multicolumn{3}{c}{Note--Parenthesis show statistical errors.}\\
\end{tabular}
\end{table}

Based on the $^{13}$CO (6--5) observations, we estimate the $^{12}$CO (6--5) line opacity at the FIR 6b and FIR 6c-B1 positions. We follow the formulation introduced in Paper I, in which under the assumption that the $^{13}$CO emission is optically thin and that $^{12}$CO and $^{13}$CO have the same excitation temperature, the $^{12}$CO to $^{13}$CO antenna temperature line ratio provide a measure of the $^{12}$CO opacity, $\tau_{12}$ (see equation 1 of Paper I). Our opacity calculations were made assuming a [$^{12}$CO]/[$^{13}$CO] abundance ratio of 50 \citep{Kahane18}. Note also that the error of the ratios does not require additional calibration uncertainties, since the observations were made with the same instrument and within the same band. In Table \ref{opacity} we show the opacities estimated for the FIR6b and FIR6c-B1 positions. These estimations are performed using the three channels around the cloud velocity. The $\tau_{12}$ values vary from $\sim$ 2 to 10. Therefore we see that the $^{12}$CO is optically thick at the cloud velocity. On the other hand, in the outflow velocity range, where $^{13}$CO was not detected, the 3$\sigma$ upper limits indicate that the emission can be optically thin: with an intensity ratio $>50$, $\tau_{12}$ is $<$0.4. Based on these results, in the calculations presented in the analysis, to determine the gas excitation and outflow properties, we assume that the $^{12}$CO (6--5) high-velocity emission is optically thin in the FIR 6 outflows, and in this way use the simple formulations for the optically thin case. Although we did not observe the $^{13}$CO (6--5) line in MMS 1-6 region, we will also assume that the $^{12}$CO (6--5) high-velocity emission from the outflows in that region is optically thin.

\subsubsection{SiO (5--4) emission}

The SiO (5--4) emission is only detected to the SW of FIR 6b, in the velocity range from -12 to 15 km s$^{-1}$. The upper panel of Figure \ref{siomap} shows the SiO (5--4) integrated emission map. The SiO emission is located towards the region that corresponds to the blue-lobe of the FIR 6c outflow. The angular resolution ($\sim$30$''$) of the SiO (5--4) observations is not sufficient to allow a clear identification of outflow structures. The lower panel of Figure \ref{siomap} shows the SiO (5--4) spectrum averaged over the area in which the emission is detected (indicated by the green polygon in the upper panel of Figure \ref{siomap}). The SiO (5--4) averaged spectrum shows a velocity range $-$12 $\leq$ V$_{LSR}$ $\leq$ $+$15 km s$^{-1}$. The lower panel of Figure \ref{siomap} also shows as comparison with the CO (6--5) spectrum averaged over the same region, obtained from a map convolved to the same angular resolution. The velocity range is similar, suggesting that the SiO (5--4) and CO (6--5) are tracing the same  outflow gas. However, while the CO profile shows a typical wing-like profile with a main peak at low velocities, the SiO is not wing-like and with a secondary peak at high velocities (V$_{LSR}$ $\sim$ $-$8 km s$^{-1}$). This behaviour, also observed in other outflows, is indicative of the uniqueness of the SiO as a shock tracer \citep[e.g., L1448: Paper I;][]{Bachiller90}. In CO (6--5) an indication of secondary peaks at high velocities is only observed in the spectra at the original angular resolution (Figure \ref{spect-65-fir6}), in fact at about the same SiO (5--4) velocity range. 

Our APEX SiO (5--4) data showed the SiO gas distribution similar to that found in previous interferometric SiO (2--1) observations by \citet{Shima09}. In both cases the SiO emission was detected only towards the region that corresponds to the blue lobe of the FIR 6c outflow. Our larger map confirms that no SiO emission is clearly detected from the FIR 6b outflow. Similar cases where CO outflows do not present an SiO counterpart have been reported in the literature \citep[see][for discussion]{Zapata06}.  

\subsection{OMC-3 MMS 1-6 outflows}

\subsubsection{CO (6--5) and CO (7--6) emission}

      \begin{figure*}
   \centering
      \includegraphics[angle=90,width=14cm]{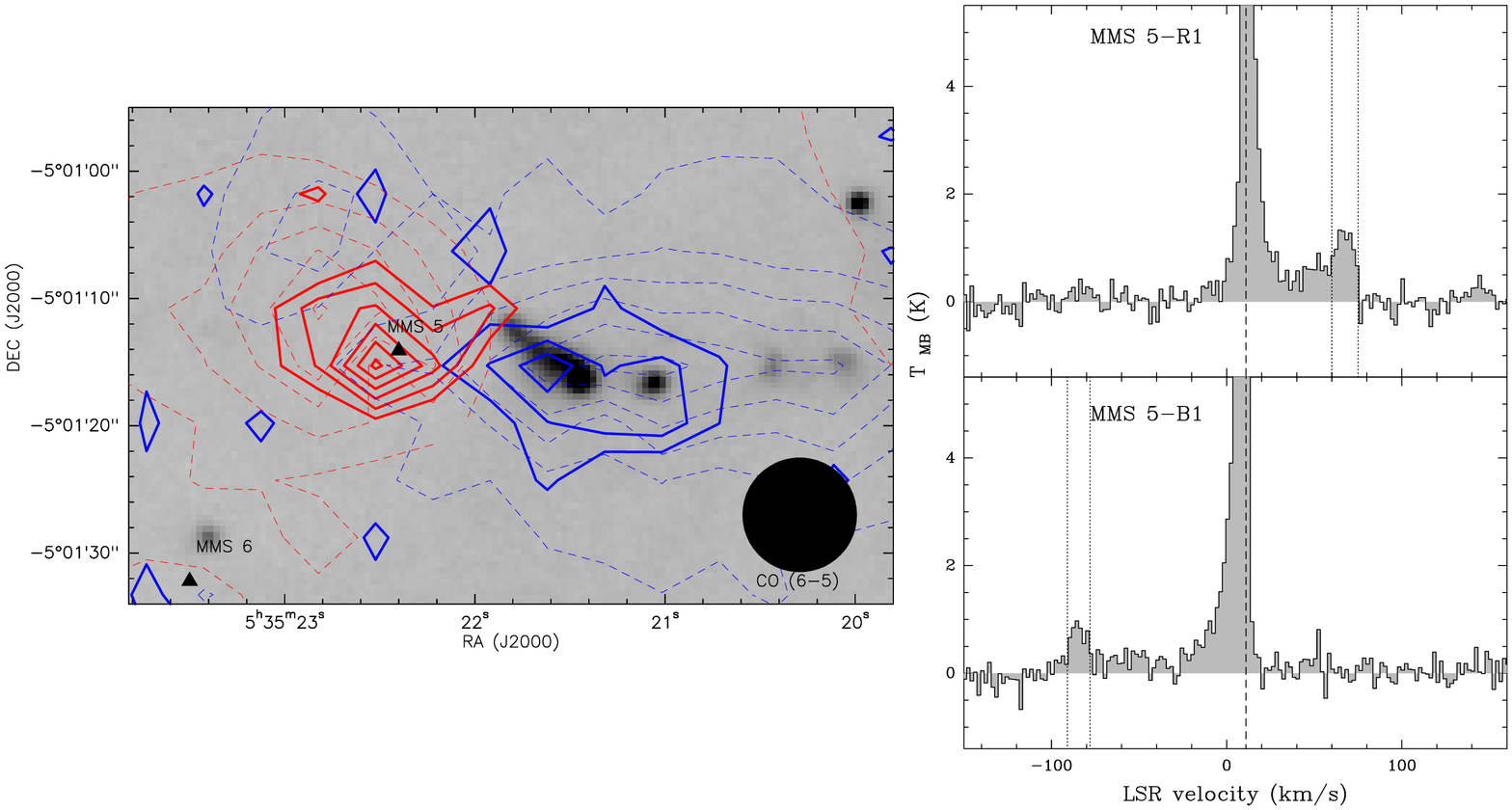}
      \caption{{\it Left:} EHV CO (6--5) emission from MMS 5 outflow. The solid blue contours show the CO (6--5) emission integrated from V$_{\rm LSR}=$-91 to -78 km s$^{-1}$ (blue-shifted EHV range), while the solid red contours show the CO (6--5) emission integrated from V$_{\rm LSR}=$ 60 to 75 km s$^{-1}$ (red-shifted EHV range). First contour is 3$\sigma$, while contour spacing in steps of $\sigma$ (2.2 K km s$^{-1}$). The dashed contours show the high-velocity CO (6--5) emission as presented in Figure \ref{mms5-co65-wings}. The background grey scale image is the NIR (K-band, 2.12$\mu$m) map taken from T08. {\it Right:} CO (6--5) spectrum taken at the B1 and R1 positions (see Table \ref{positions}). The vertical dot-dashed lines show the boundaries of the EHV range, while the dashed line indicates the cloud velocity.
              }
         \label{EHV-mms5}
   \end{figure*}

Figure \ref{mms5-co65-wings} shows the region covered by our CO (6--5) and (7--6) maps. We have clearly identified CO (6--5) outflow emission originating from the sources MMS 5 and MMS 2. We identified four outflow peaks, listed in Table \ref{positions}. These two outflows were previously reported by T08. However the compact outflow detected in CO (3--2) by \citet{Taka12} emanating from MMS 6 was not detected in CO (6--5). The non-detection of the latter outflow is possibly due to beam dilution, since even with Nobeyama interferometric CO (1--0) observations by \citet{Taka09}, with an angular resolution of several arc-seconds, the compact outflow was not detected \citep[see also discussion by][]{Taka12}. No clear evidence of outflow activity towards the remaining MMS objects was found. The outflow emanating from MMS 5 seems to be orientated along the East-West direction, while for the outflow from MMS 2 the observations can not clearly disentangle a tendency in the outflow direction. We also notice two blobs of blue-shifted emission in the far side of the putative MMS 2 blue-shifted outflow lobe, which coincide with similar structures detected in CO (3--2). Since T08 and \citet{Will03} could not unambiguously relate these structures with with MMS2 (or the other nearby millimeter sources), for consistency with the later physical parameters comparisons, we did not take these structures into account in the following calculations. The most intense CO (6--5) emission is related to the MMS 5 outflow, which shows an extended blue lobe, while the red lobe shows a more compact structure. 

A remarkable result from our CO (6--5) observations is the detection of EHV emission from the MMS 5 outflow. The EHV range is defined as V$_{\rm LSR}=$-95 to -80 km s$^{-1}$ for the blue-shifted lobe and V$_{\rm LSR}=$60 to 70 km s$^{-1}$ for the red-shifted lobe. In Figure \ref{EHV-mms5} we show the velocity integrated emission in the EHV range, overlaid on the near infrared (NIR) emission (K-band, 2.12$\mu$m, taken from T08), and the spectra taken at the position of the peak intensity of the EHV structures. The integrated intensity maps show that the EHV emission is located within a radius of $\sim$ 10$''$ from MMS 5. The blue-shifted EHV emission coincide with the inner collimated NIR structure. The CO (6--5) spectra clearly show secondary peaks related to these EHV features, with the red-shifted EHV component stronger than the blue-shifted one. The compact morphology and the characteristic line profile suggest that these components are the equivalent of the so-called EHV bullets found in class 0 low-mass outflows \citep[e.g. L1448-mm:][paper I;]{Bachiller90}.  Although evidence for EHV emission was not found in  the CO (3--2) spectra observed by T08, recent ALMA observations of CO (2--1) and SiO (5--4) have revealed the EHV emission from MMS 5 \citep{Matsushita18}.

\begin{figure}
   \centering 
\includegraphics[angle=0,width=7cm,angle=90]{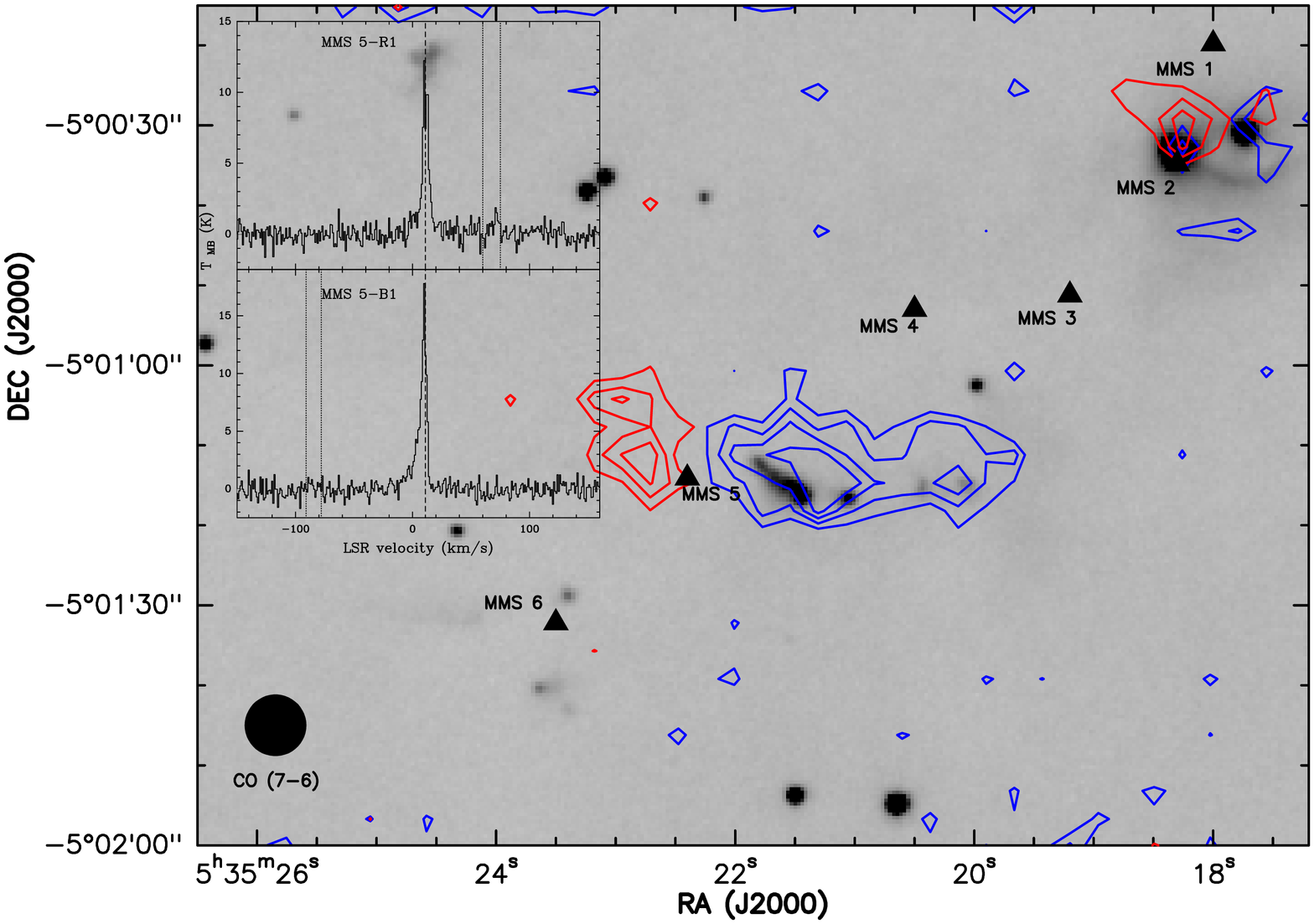}
      \caption{The CO (7--6) high-velocity emission map centered on MMS 5. Blue and red contours show the blue- and red-shifted high-velocity emission (integrated in the same velocity ranges as in Figure 2), respectively. Contour spacing is 1$\sigma$, with 3$\sigma$ as the first contour ($\sigma$=7.4 K km s$^{-1}$ for blue-shifted emission and $\sigma=$6.7 K km s$^{-1}$ for red-shifted emission). The background grey scale image as in Figure 4. Filled triangles as in Figure 2. The inlay shows the CO (7--6) spectrum taken at the B1 and R1 positions, with the vertical lines as in Figure \ref{EHV-mms5}. 
              }
         \label{mms5-76-wings}
   \end{figure}

The CO (7--6) high-velocity emission was detected in the outflows associated with MMS 5 and MMS 2 (see Figure \ref{mms5-76-wings}). The outflow lobes from MMS 5 in the CO (7--6) emission show a similar structure to those traced in the CO (6--5) emission. An EHV component in the CO (7--6) line is barely noticed at 3$\sigma$ only at the R1 position. In the case of the MMS 2 outflow, the CO (7--6) high-velocity emission is coming mainly from the red-shifted lobe, with the blue-shifted lobe only detected at 4$\sigma$ level. 

\section{Analysis}

\subsection{Population diagram}

\begin{table}
\caption{CO integrated intensities at selected positions$^a$.} 
\label{table:b1r1}      
\centering          
\begin{tabular}{l c c c c}     
\hline\hline       
Position & Velocity (km/s) & \multicolumn{3}{c}{$\int$ $T_{MB}$$d\varv$($\sigma$) (K km s$^{-1}$)$^b$}\\
\cline{3-5}
 &($\varv_{min}$,$\varv_{max}$) & 3--2 & 6--5 & 7--6    \\
\hline                    
FIR 6b-B1	&0,8	&32(2)	&17(1)	&$<$4$^c$\\
FIR 6b-R1	&14,30	&22(1)	&27(1)	&20(4)\\
FIR 6c-B1	&-16,8	&23(1)	&27(2)	&20(5)\\
FIR 6c-B2	&-10,8	&33(1)	&39(1)	&20(5)\\
MMS 5-B1	&-25,7	&36(2)&61(1)&42(4)\\
EHV		&-95,-75	&$<$10$^c$	&11(1)	&3(3)\\
MMS 5-R1	&16,31	&6(2)	&16(1)	&8(2)\\
EHV		&60,75	&$<$7$^c$&14(1)	&8(2)\\
MMS 2-B1         &-12,7  &16(2)      &18(1)      &12(3)\\
MMS 2-R1		&16,30	&9(1)	&11(1)	&8(3)\\
	&31,50	&$<$8$^c$		&7(1)	&$<$3$^c$\\
\hline
\hline       
\multicolumn{5}{l}{{a) Data convolved to 26$''$ ASTE beam (i.e. CO (3--2) transition).}}\\
\multicolumn{5}{l}{{b) Integrated between $\varv_{min}$,$\varv_{max}$. Statistical errors in parenthesis.}}\\ 
\multicolumn{5}{l}{c) 3$\sigma$ upper limit.}\\          
\end{tabular}
\end{table}

Under the assumption that the gas is in LTE and that the $^{12}$CO emission is optically thin, we use the Boltzmann relation for the level populations to estimate the rotational temperature ($T_{\rm rot}$) and the total column density ($N$) at selected positions within the outflow lobes, following the formulation presented in Paper I. The calculations were made for the emission from the outflow peaks reported in Table 3 (see also Figures \ref{fir6-co65-wings} and \ref{mms5-co65-wings}). The CO (3--2) data taken with ASTE by T08 were added to our APEX data. We convolved our CO (6--5) and (7--6) observations to the angular resolution of the CO (3--2) observations (i.e. 26$''$). Table \ref{table:b1r1} shows the CO (3--2, 6--5, 7--6) integrated intensities of the selected positions, convolved to 26$''$ resolution. The emission is integrated between the outflow's wing limits, $\varv_{min}$ and $\varv_{max}$, which we assume to be optically thin (see sect. 4). In the calculation we included the calibration uncertainties of each line, whose value is 20\% for all the cases (including the ASTE data; see T08). For future references, we report separately in Table \ref{table:b1r1} the intensities for the EHV range, although we do not provide calculations for this component, given the mostly non-detection of CO (3--2) and CO (7--6) lines. 

Table \ref{table:rotdigram} shows the results, $T_{\rm rot}$ and $N$, obtained from the rotational diagram analysis. To within the errors, most of the positions present a rotational temperature in the range of 60-70 K, with the exception of the FIR 6b-B1 position, which shows a lower temperature of $\sim$ 30 K. In terms of the column density, higher values are found in the OMC-2 FIR 6 positions.


\begin{table}
\caption{Rotational diagram} 
\label{table:rotdigram}      
\centering          
\begin{tabular}{l l l}     
\hline\hline       
Position & $T_{\rm rot}$(K)   & $N$ (10$^{16}$ cm$^{-2}$)  \\
\hline
\hline                    
FIR 6b-B1 & 37$\pm$9 & 1.5$\pm$0.8 \\
FIR 6b-R1 & 69$\pm$4 & 1.2$\pm$0.1 \\
FIR 6c-B1 & 67$\pm$4 & 1.2$\pm$0.1 \\  
FIR 6c-B2 & 60$\pm$10 & 1.7$\pm$0.5 \\
MMS 5-B1 & 83$\pm$12 & 2.2$\pm$0.4 \\
MMS 5-R1 & 95$\pm$65 & 0.4$\pm$0.3\\  
MMS 2-R1 & 68$\pm$5 & 0.48$\pm$0.05\\ 
MMS 2-B1 & 63$\pm$4 & 0.83$\pm$0.09 \\   
\hline
\hline      
\multicolumn{3}{l}{We have used integrated intensities from Table \ref{table:b1r1}.}\\
\multicolumn{3}{l}{Statistical and calibration errors are taken into account.}\\ 
\end{tabular}
\end{table}  
  
\subsection{Large Velocity Gradient}

Radiative transfer calculations with RADEX (van der Tak et al. 2007) were made in the Large Velocity Gradient (LVG) approximation and plane-parallel geometry. The molecular data was retrieved from the LAMDA database \footnote{http://www.strw.leidenuniv.nl/~moldata/} \citep{LAMDA} and the collisional rate coefficients are adopted from Yang et al. (2010). 


The RADEX calculations are used to reproduce simultaneously the observed CO (3--2)/CO (6--5), CO (3--2)/CO (7--6), and CO (6--5)/CO (7--6) integrated intensity ratios at different positions within the outflow lobes. The intensities are taken from Table \ref{table:b1r1}, and we added the calibration uncertainties already reported (note, however, that the ratios with the CHAMP+ lines themselves do not require additional calibration errors, since they were observed simultaneously).

\begin{figure*}
   \centering
   \includegraphics[angle=-90,width=15cm]{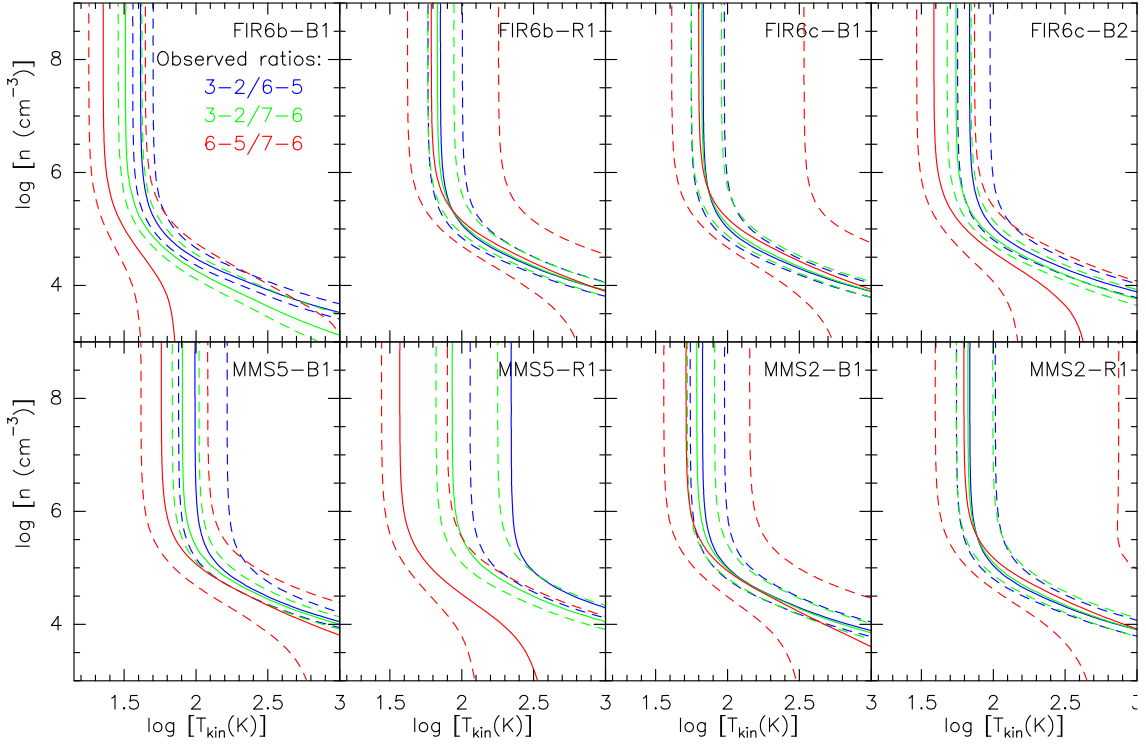}
      \caption{{\bf Kinetic} temperature (T$_{\rm kin}$) versus H$_2$ density (n) plots from LVG calculations. The curves show the observed (3--2)/(6--5), (3--2)/(7--6), and (6--5)/(7--6) integrated intensity ratios (computed from Table \ref{table:b1r1}). The dashed lines represent the errors in the line ratios. The positions are indicated by the labels in each panel.  
              }
         \label{omc-nT}
   \end{figure*}

In Figure \ref{omc-nT}, kinetic temperature (T$_{\rm kin}$) vs. H$_2$ density (n) plots are shown for each outflow position. In most of the cases only lower limits to kinetic temperature and H$_2$ density can be provided, with values of T$_{\rm kin} >$ 30--50 K and n$>$ 10$^4$ cm$^{-3}$. Only in the case of MMS 5-R1 the three ratios are able to set more stringent constraints to the density and a higher kinetic temperature lower limit: n$\approx$10$^4$--10$^5$ cm$^{-3}$, T$_{\rm kin} >$ 200 K. The later values indicate that the outflow driven by MMS 5 present physical conditions similar to those obtained in the low-mass class 0 outflows from L1448-mm and HH211 (see Paper I).

\subsection{Outflow Parameters}

\begin{table*}
\small
\caption{Outflow properties derived from CO (6--5).}             
\label{outparam}      
\centering          
\begin{tabular}{l c c c c c c c}     
\hline\hline       
Velocity & $\delta \varv_{\rm max}$$^{(1)}$ & $R_{\rm max}$$^{(2)}$ & $t_{\rm d}$$^{(3)}$ &  M(H$_2$)$^{(4)}$& $F_{\rm m}$$^{(5)}$ & $E_{\rm k}$$^{(6)}$ &$L_{\rm mec}$$^{(7)}$\\
& & (10$^{-2}$) & (10$^3$) & (10$^{-3}$) & (10$^{-5}$) & (10$^{43}$) &\\
range    & (km s$^{-1}$)& (pc) & (yr) & (M$_{\odot}$)& (M$_{\odot}$ km s$^{-1}$ yr$^{-1}$) & (erg) & (L$_{\odot}$)\\
\hline
\multicolumn{8}{c}{FIR 6b}\\
\hline                    
Blue	&8     &3 &4 & 6     	&2 &0.7  & 0.02\\
Red	&19    &4 &2 & 3     	&3 &2  & 0.07\\
Total   &      &     &      & 9      &5 &2.7  & 0.09\\
\hline
\multicolumn{8}{c}{FIR 6c}\\
\hline
Blue	&26	&4 &1 & 5	&13 &7 & 0.4\\
\hline
\multicolumn{8}{c}{MMS 5}\\                 
\hline
Blue	&106	&6 &0.6 & 9    	&230 &200  & 28\\
Red	&59	&4 &0.6 & 5    	&70 &30  & 5\\
Total   &       &     &      & 14  &300 &230  & 33\\
\hline
\multicolumn{8}{c}{MMS 2}\\
\hline
Blue	&23	&4 &2 & 1    	&2 &1 & 0.06\\
Red	&19	&4 &2 & 0.4    	&0.5 &0.3 & 0.01\\
Total   &       &    &    & 1.4         &2.5 &1.3 & 0.07\\
\hline
\hline
\multicolumn{8}{l}{$^{(1)}$Maximum relative velocity of the lobe. $^{(2)}$ Maximum extension of the lobe. $^{(3)}$ Dynamical time.  }\\
\multicolumn{8}{l}{$^{(4)}$H$_2$ mass. $^{(5)}$ Mechanical force. $^{(6)}$ Kinetic energy. $^{(7)}$ Mechanical luminosity}\\
\end{tabular}
\end{table*}

Using the velocity integrated CO (6--5) high-velocity maps (pixels with signal-to-noise ratio above 3), we calculate the H$_2$ mass of the outflow gas, following the relations presented in Paper I between the total CO column density ($N$) and the velocity integrated emission, under the assumption of optically thin $^{12}$CO (shown to be the case, based on the opacity upper limit determined in Sect. 4 for the high-velocity gas at some outflow positions). The high-velocity maps are produced by integrating the emission between the minimun and maximum velocity of each outflow lobe, following the $\varv_{\rm min}$,$\varv_{\rm max}$ values reported in Table \ref{positions}. In the calculations we used the temperatures obtained from the rotational diagram analysis (Table \ref{table:rotdigram}), the standard ISM value for the relative $^{12}$CO abundance, X(H$_2$/CO) $\sim$ 10$^{4}$ \citep[e.g.,][]{Wilson92}, and a distance to the source of 400 pc (which is close to the average of the distance range of Orion A 3-D structure). As in Paper I, we point out that considering uncertainties such as data calibration, definition of the velocity boundary limits for the integrated emission, and opacity assumptions, the mass determinations are usually accurate within a factor of 2-3 \citep[see, e.g.,][]{Cabrit90}. 

The mass detection limit of our CO (6--5) observations, assuming a $T_{ex}$ of 60 K (following Table \ref{table:rotdigram}) and the r.m.s. at 1 km s$^{-1}$, is $\sim$ 1$\times$10$^{-5}$ M$_{\odot}$ inside a beam of 9$''$. For each outflow the H$_2$ mass is reported in Table \ref{outparam}. The total H$_2$ masses range from $\sim$ 1$\times$10$^{-2}$  M$_{\odot}$ to 1$\times$10$^{-3}$  M$_{\odot}$. We note that our mass estimates based on the CO (6--5) emission are lower by a factor of 4 with respect to those reported in Table 3 of T08 for MMS 5, MMS 2, and FIR 6c (blue-lobe), while it is lower by a factor of 9 for FIR 6b outflow. The higher discrepancy between the mass estimates from CO (3--2) and CO (6--5) in FIR 6b outflow may be mostly due to the fact that our CO (6--5) map misses the upper half of the red-shifted lobe, a region that was included in the CO (3--2) mass estimation from T08. By recomputing the mass from the CO (3--2) emission in an area equivalent to our CO (6--5) map, the difference is reduced to a factor of 3. The lower masses obtained with respect to the CO (3--2) observations may be due the typical mass estimation uncertainties we have already mentioned above; however, by looking at the maps it is clear in some cases (such as the MMS 2 outflow) that the emitting region in both lines is not exactly the same, with the CO (6--5) emission showing systematically more compact structures.

We also determine other parameters such as dynamical time scale ($t_{\rm d}$), mass outflow rate ($\dot{M}$), mechanical force ($F_{\rm m}$), kinetic energy ($E_{\rm k}$), and mechanical luminosity ($L_{\rm mech}$). These quantities are defined as follows: $t_{\rm d}$=R/V, $\dot{M}$=M/$t_{\rm d}$, $P$=M$\times$$\delta \varv_{\rm max}$, $F_{\rm m}$=M$\times$V/$t_{\rm d}$, E$_{\rm k}$=M$\times$$\delta \varv_{\rm max}^2$/2, $L_{\rm mech}$=E$_{\rm k}$/$t_{\rm d}$ \citep[e.g.][]{Beuther02}. In the above expressions, R=$R_{\rm max}$/sin $i$ and V=$\delta \varv_{\rm max}$/cos $i$, where $R_{\rm max}$ and $\delta \varv_{\rm max}$ are the maximum extension and maximum relative velocity of the lobes, and $i$ the angle of the lobe with respect to the plane of the sky. To compare with T08 we assumed $i=$ 45 degrees for all outflows. The table \ref{outparam} shows our results. Here we also point out that considering the uncertainties in the mass estimation plus the geometry assumptions mentioned above, the kinematic parameters should be accurate within a factor of 10-30 \citep[see, e.g., discussion in][]{Cabrit90}. Within the uncertainties mentioned above our results are consistent with T08.

\section{Discussion}

\subsection{Comparison with other mid-J CO observations}

The outflows properties determined for our set of intermediate-mass outflows are not very different to the values found by other authors in low-mass outflows \citep[e.g., Paper I;][]{Curtis10}. In particular, $F_{\rm m}$ and $E_{\rm k}$ values for the intermediate-mass outflow MMS 5 (3$\times$10$^{-3}$ M$_{\odot}$ km s$^{-1}$ yr$^{-1}$ and 2$\times$10$^{45}$ erg, respectively) are similar, within the typical errors, to those found in Paper I for the low-mass class 0 outflow L1448-mm ($\sim$ 8$\times$10$^{-4}$  M$_{\odot}$ km s$^{-1}$ yr$^{-1}$ and 2$\times$10$^{45}$ erg, respectively). We therefore conclude that the intermediate-mass outflow MMS 5 is as energetic as its low-mass counterpart in L1448-mm. However, we point out that this may be only a particular case, since previous studies toward statistical significant samples have shown a trend of increasing energy and mechanical luminosity as the protostellar masses increase \citep[e.g.,][]{Beuther02}.  

\citet{Kempen15} presented CO (6--5) observations of a sample of 6 intermediate-mass protostars. They determined outflow parameters employing methods similar to ours, allowing for a direct comparison. In general our results are in agreement with those of \citet{Kempen15}. In  particular, the outflow force is within the same range, except for the case of NGC2071. More discrepant are the outflow masses which are higher for most of the sources in the \citet{Kempen15} sample. The latter is likely due to the larger size of the outflow lobes in that sample, which in turn may be an effect of the age. Given that we obtained a similar outflow force range as found for the \citet{Kempen15} sources that span a similar range of bolometric luminosities of the central objects, the correlation between the outflow force and the bolometric luminosity holds for both our and their sample. 


\subsection{The EHV collimated jet from MMS 5}

Figures \ref{mms5-76-wings} and \ref{EHV-mms5} show the CO (6--5) and CO (7--6) high-velocity and EHV emission, overlaid with the NIR emission from the MMS 5 region. The main NIR features in this region are an extended monopolar structure and three knots to the west of it. These features are coincident with the CO (6--5) and CO (7--6) blue-shifted emission. On the other hand, no NIR structure is coincident with the red-shifted CO lobe, likely because of the known affect of extinction from the foreground gas in the red-shifted lobes (counter-jet) \citep[see e.g.,][]{Noriega12}. The NIR monopolar structure is indeed well collimated (size $\sim 8.7''\times1.7'' $, hence a collimation factor $\sim$ 5), which suggests the jet nature of the NIR emission. Since the knots to the west are not well aligned with this monopolar structure, it is possible that the jet from MMS 5 is precessing. As noted in Figure \ref{EHV-mms5}, the EHV CO(6--5) blue-shifted emission seems to be coincident with the NIR collimated monopolar structure and the closest, brightest, NIR knot to the west of it. Therefore, the positional correlation with this NIR structure and knot may suggest the jet nature of the CO EHV emission. The higher angular resolution ALMA data reveals a collimated structure of the EHV gas in the CO (2--1) and SiO (5--4) transitions \citep{Matsushita18}. The similar line profile and velocity range of the ALMA CO observations indicates that our mid-J CO observations are tracing the same outflow from MMS 5. However, the latter authors have found that the EHV gas and the NIR jet are not positionally coincident, suggesting that the H$_2$ knots may not be related to the CO and SiO jets revealed by their data.

To our knowledge Cep-E and NGC2071 are the only other confirmed cases of EHV emission from intermediate-mass outflows \citep{Lefloch15,GR12,Chernin92,Hatchell99}, so the discovery of EHV gas from the MMS 5 outflow is an important addition to this kind of objects that would help to understand the jet phenomenon within the context of star-formation theories, in particular for the intermediate-mass case. Taking into account only the EHV emission and assuming that our CO (6--5) data resolved the major axis of the blue-shifted lobe, a dynamical time scale of $\sim$ 150 yr is estimated for this kinematic structure, which is within the range measured in other very young IM outflows \citep[e.g.,][]{Zapata06}. This result highlight the relevance of the mid-J CO lines to search for outflows in their early phase of evolution.

\section{Summary and conclusions}

The main results of this study can be summarized in this way:

   \begin{itemize}
      \item The CO (6--5) (and in some cases the CO (7--6)) emission in OMC-2 FIR 6 and OMC-3 MMS 1-6 was found tracing outflows related to class 0 type intermediate-mass objects within those regions.

      \item The CO (6--5) line profiles particularly highlight the secondary peaks, likely related to shock structures.

      \item Extremely High Velocity CO (6--5) emission was detected from the MMS 5 outflow in the OMC-3 region. The CO (7--6) transitions was detected at the 3$\sigma$ level from the red lobe of the same outflow.

      \item By a comparison with previous observations, we found that at positions close to the central objects the CO (6--5) and (7--6) emission is stronger than the low-$J$ CO (3--2) emission. This behavior is similar to what we found in low-mass outflows (L1448 and HH211). 

       \item The kinematics and physical conditions of MMS 5 outflow suggest that it is the counterpart of the young class 0 low-mass outflows. 

   \end{itemize}

\begin{acknowledgements}
      We wish to thank all the APEX staff in Chile for their enthusiastic help during these observations. AIGR was supported through an stipend from the International Max-Planck Research School for Astronomy and Astrophysics at the universities of Bonn and Cologne, and is currently supported by CONACYT under the program C\'atedras CONACYT para j\'ovenes investigadores. A. Gusdorf acknowledges support by the grant ANR-09-BLAN-0231-01 from the French Agence Nationale de la Recherche as part of the SCHISM project.
\end{acknowledgements}

\bibliography{bibliog}

\end{document}